\documentclass[%
 aip,
 amsmath,amssymb,
 reprint,%
]{revtex4-1}

\usepackage{graphicx}% Include figure files
\usepackage{dcolumn}% Align table columns on decimal point
\usepackage{bm}% bold math
\usepackage[utf8]{inputenc}
\usepackage[T1]{fontenc}
\usepackage{mathptmx}

\usepackage{hyperref}
\usepackage{url}
\usepackage{epstopdf}
\usepackage{xfrac}
\usepackage{multirow}
\usepackage{longtable}

\newcommand{\fig}{Fig.}

\newcommand{\SW}{SW}
\newcommand{\centre}{\omega_{0}}
\newcommand{\width}{\gamma}

\begin{document}

\preprint{AIP/123-QED}

\title[Backfolded phonons in metal-oxide superlattices]{Backfolded acoustic phonons in metal-oxide superlattices}
% Force line breaks with \\

\author{F. Lyzwa}
\affiliation{University of Fribourg, Department of Physics and Fribourg Center for Nanomaterials, Chemin du Mus\'{e}e 3, CH-1700 Fribourg, Switzerland}

\author{A. Chan}
\affiliation{The MacDiarmid Institute for Advance Materials and Nanotechnology and The Dodd Walls Centre for Quantum and Photonic Technologies, New Zealand}
\affiliation{School of Chemical Sciences, The University of Auckland, Auckland, New Zealand}

\author{J. Khmaladze}
\affiliation{University of Fribourg, Department of Physics and Fribourg Center for Nanomaterials, Chemin du Mus\'{e}e 3, CH-1700 Fribourg, Switzerland}

\author{K. F\"{u}rsich}
\affiliation{Max-Planck-Institut f\"{u}r Festk\"{o}rperforschung, Heisenbergstrasse 1, 70569 Stuttgart, Germany}

%\affiliation{Robinson Research Institute, Victoria University, P.O. Box 600, Wellington, New Zealand}
%\affiliation{The MacDiarmid Institute for Advance Materials and Nanotechnology and The Dodd Walls Centre for Quantum and Photonic Technologies, New Zealand}
%\affiliation{School of Chemical Sciences, The University of Auckland, Auckland, New Zealand}
%	\affiliation{Department of Physics and School of Chemical Sciences, XXXX, The Photon Factory, The University of Auckland, 38 Princes St, Auckland, New Zealand}
%	\affiliation{University of Fribourg, Department of Physics and Fribourg Center for Nanomaterials, Chemin du Mus\'{e}e 3, CH-1700 Fribourg, Switzerland}

\author{B. Keimer}
\affiliation{Max-Planck-Institut f\"{u}r Festk\"{o}rperforschung, Heisenbergstrasse 1, 70569 Stuttgart, Germany}

\author{C. Bernhard}
\affiliation{University of Fribourg, Department of Physics and Fribourg Center for Nanomaterials, Chemin du Mus\'{e}e 3, CH-1700 Fribourg, Switzerland}

\author{M. Minola}
\affiliation{Max-Planck-Institut f\"{u}r Festk\"{o}rperforschung, Heisenbergstrasse 1, 70569 Stuttgart, Germany}

\author{B.P.P. Mallett}
\affiliation{The MacDiarmid Institute for Advance Materials and Nanotechnology and The Dodd Walls Centre for Quantum and Photonic Technologies, New Zealand}
\affiliation{School of Chemical Sciences, The University of Auckland, Auckland, New Zealand}
\email{benjamin.mallett@gmail.com}

\date{\today}% It is always \today, today,

\begin{abstract}
We report the observation of low-frequency modes in the Raman spectra of thin-film superlattices of the high-temperature superconductor YBa$ _{2} $Cu$ _{3} $O$ _{7-\delta} $ and various manganite perovskites. Our study shows that these modes are caused by the backfolding of acoustic phonons due to the additional periodicity introduced by the superlattice. Such modes were previously only observed for ultra-pure semiconductor superlattices. They can be used to determine the bilayer thickness of the superlattice and its speed of sound. Moreover, we use the spatial resolution of Raman microscopy to map the film thickness inhomogeneity across a sample, making these modes a useful tool to characterize thin-film superlattices.
\end{abstract}

\maketitle

%\section{Introduciton}

Thin-film multilayers of metal-oxides host a wide range of emergent, tunable and potentially useful properties \cite{chakhalian2007, reyren2007, bibes2011, hwang2012, driza2012, rogdakis2012, chakhalian2014, lorenz2016, mallett2016pyp, keunecke2019} beyond those observed in multilayers of more conventional semi-conductors such as Si or GaAs \cite{smith1990, mannhart2010}. %and grahn1995, R. Chau, B. Doyle, S. Datta, J. Kavalieros, and K. Zhang, Nature Mater. 810 (2007), pp. 810–812
This is owed to the various types of magnetic, charge, ferroelectric and superconducting orders that the constituent metal-oxides layers can host \cite{tokura2006, keimer2015}. 
The novel properties of thin-film multilayers and superlattices result from interactions across the interface, such as spin and orbital reconstruction, charge-transfer and phonon-coupling \cite{bibes2011, hwang2012, driza2012, chakhalian2014}, as well as some hitherto unidentified mechanisms \cite{mallett2016pyp, perret2018}. 
As such, multilayers of metal-oxides constitute a fertile playground both to discover interesting physics and to tailor functionalities that could shape future electronics.

Besides the scientific challenge of understanding the properties of metal-oxide superlattices, there is the persistent technical challenge of maintaining the quality of crystal structure, interfaces and layer thickness in such samples \cite{norton2004}. Growth techniques for such samples include\cite{norton2004} pulsed laser deposition (PLD) \cite{dijkkamp1987}, molecular beam epitaxy (MBE) \cite{schlom1988} and sputtering \cite{rao1996}.

Here we report on the observation of low-frequency Raman modes in metal-oxide superlattices, with a focus on superlattices of the high-temperature superconductor YBa$ _{2} $Cu$ _{3} $O$ _{7} $ and $ R $MnO$ _{3} $ manganites, which are exemplary multilayers for displaying the effects described above \cite{chakhalian2014, mallett2016pyp, perret2018}. The low-energy modes are optical-phonons which arise in superlattices due to the backfolding of the $ c $-axis acoustic phonon branch and, to the best of our knowledge, are detected for the first time in metal-oxide heterostructures.
Until now the observation of backfolded acoustic phonon modes has only been reported in superlattices of high-quality MBE-grown semiconductor superlattices, such as GaAs/AlAs \cite{colvard1980, cardona1989, rouhani1991, sapriel1991}. The occurence and the properties of these modes was fully explained and described within a comprehensive theoretical framework \cite{colvard1985, cardona1989, dharma-wardana1993}.
The modes can thus be used to characterize the quality and properties of the superlattice, such as the speed of sound and/or the bilayer thickness. Given the fast acquisition time of the measurement and the potential for micrometer spatial resolution, such modes can be a useful diagnostic for characterizing superlattices of given materials. % in principle could be an in-situ measurement - though this would present several technical challenges.

%\section{Methods}
%
%\subsection{Sample growth}
In particular, in the present work we focus on epitaxial superlattices of the cuprate superconductor YBa$ _{2} $Cu$ _{3} $O$ _{7} $  (YBCO) and manganite $ R $MnO$ _{3} $, where $ R =$ Pr$ _{0.5} $La$ _{0.2} $Ca$ _{0.3} $ (PLCMO), Nd$ _{1-x} $(Ca$ _{1-y} $Sr$ _{y} $)$ _{x} $ (NCSMO) or La$ _{1-x} $Sr$ _{x} $ (LSMO). Details of the samples can be found in the supplementary information. The samples are  grown by PLD on La$ _{0.3} $Sr$ _{0.7} $Al$ _{0.65} $Ta$ _{0.35} $O$ _{3} $ (LSAT) substrates that are (0 0 1)-oriented (commercially available from Crystec) following the process described in Refs.~\cite{malik2012, perret2018}. The superlattices involve ten repetitions of cuprate-manganite layers, whose thickness we denote using the following scheme; PLCMO(10~nm)/YBCO(7~nm) translates to a 7~nm thick YBCO layer grown on top of a 10~nm thick PLCMO layer. The topmost layer is the manganite followed by a 2~nm thick LaAlO$ _{3} $ capping layer to protect the film surface from degradation.
We performed \textit{ex situ} studies with x-ray diffraction, x-ray reflectivity, and polarized neutron reflectivity to measure the layer thickness, uniformity, and interface roughness. Representative results can be found in the supplementary materials of Refs.~\cite{mallett2016pyp, perret2018}. These show our samples are of high-quality, with a small interface roughness ($ \sim 0.5 $~nm) that tends to increase with additional cuprate/manganite layers and minimal chemical diffusion across the interface.

%% Raw data - setting the scene %%
\begin{figure*}
	\includegraphics[height=5.5cm]{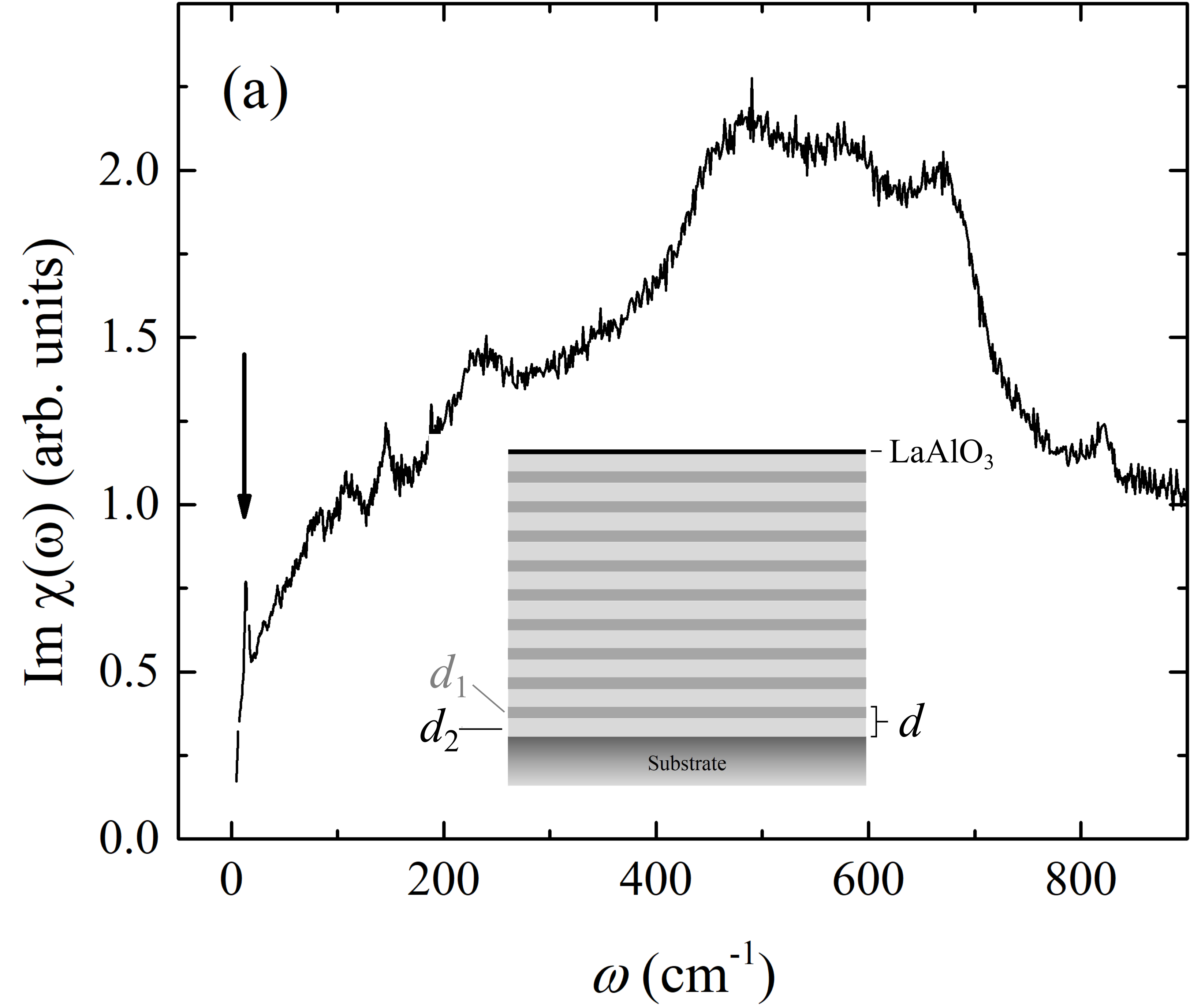}
	\includegraphics[height=5.5cm]{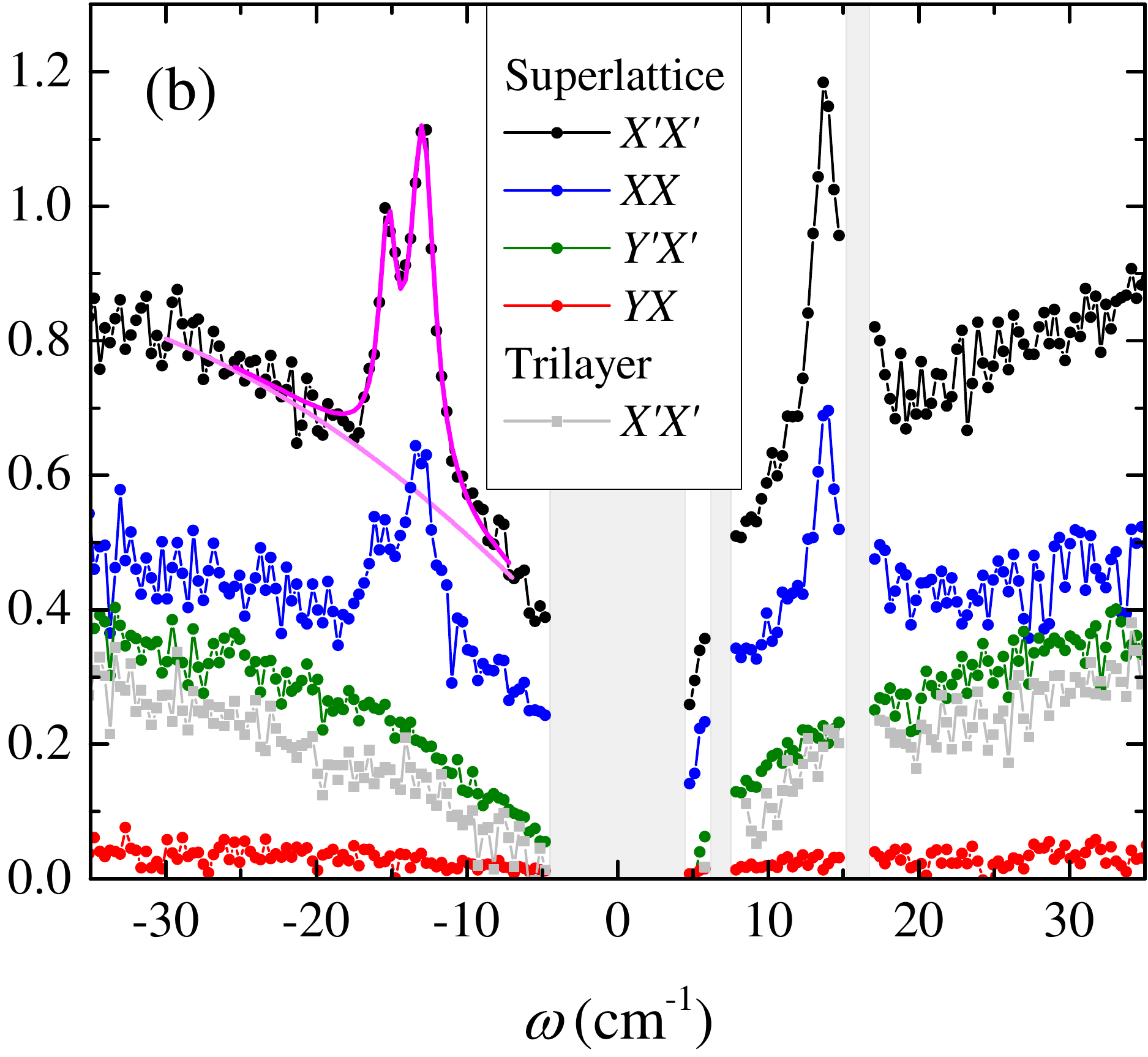}
	\caption{(a) Spectrum of a PLCMO(10nm)-YBCO(7nm) superlattice, illustrated bottom center, in $z(X'X')z$ scattering geometry showing the comparatively intense and narrow line-width of the low-frequency modes compared with the phonon modes above 70~cm$ ^{-1} $. (b) Polarization analysis of the low frequency modes. The magenta line for the $ X'X' $ superlattice spectrum shows a fit to the data with the modelled background as the light magenta line. They grey data (offset by -0.5 for clarity) are for a trilayer sample of the same composition which indicates that the low energy modes are inherent to the superlattice samples. }
	\label{fig:raw}
\end{figure*}

%\subsection{Raman measurements}
The Raman spectra were recorded with a Jobin-Yvon LabRam HR800 spectrometer using the $ 632.8 $~nm excitation line of a HeNe laser \cite{hepting2014}. The measurements were carried out in full back scattering with geometry indicated by Porto's notation. For example, $ z(Y'X')\underline{z} $ indicates backscattering with incident polarization 45$ ^{\circ} $ to the Mn-O nearest-neighbour bond with the cross-polarised scattered light measured. We find spurious reflections in our spectrometer that lead to artefacts around $ 7 $~cm$ ^{-1} $ and $ 15.8 $~cm$ ^{-1} $ - these spectral regions are removed from the reported spectra for clarity. Unless noted otherwise, the spectra shown were recorded at room temperature.
% The 1800 lines/mm gratings gives a spectral resolution of $ 0.3 $~cm$ ^{-1} $. The laser was focused with a $ \times 100 $ long-working distance objective lens with a short depth of focus, NA$ =0.6 $, which was positioned with an accuracy of $ 0.5 $~$ \mu $m such that the focus is centered on the film \cite{hepting2015}. In order to further optimize the multilayer signal, a 50~$ \mu $m confocal hole along the scattered light path to minimize the signal from the LSAT substrate. The residual substrate contribution was subtracted from the spectra using reference measurements for which the beam focus was moved into the substrate.
%Two volume Bragg-grating filters allow us to efficiently reject Rayleigh-scattering facilitating data collection down to $ 4.5 $~cm$ ^{-1} $ on either side of the elastic scattering from the laser. However, we find that spurious reflections in our spectrometer lead to artefacts around $ 7 $~cm$ ^{-1} $ and $ 15.8 $~cm$ ^{-1} $ - these spectral regions are removed from the reported spectra for clarity. 
%Laser heating effects were minimized by keeping the laser power below 1~mW. %, which results in a temperature uncertainty of less than 5~K \cite{hepting2014, hepting2015}.
All reported spectra have been divided by the Bose thermal factors to obtain the imaginary part of the Raman scattering susceptibility,  Im$ \chi(\omega) $. Further details can be found in the supplementary material.

%\section{Results}
%\subsection{Exemplar case}

To set the scene, \fig~\ref{fig:raw}(a) shows the Raman Stokes signal from a PLCMO(10nm)/YBCO(7nm) superlattice (sketched) collected at room-temperature in $ z(X'X')\underline{z} $ geometry over a wide spectral range.% In this overview spectrum, we combine data taken with 1800 lines/mm gratings and 600 lines/mm gratings. 
The spectrum reveals multiple broad overlapping phonon modes above 50~cm$ ^{-1} $ arising from both YBCO, manganite and the interaction between them. A rich spectral fingerprint is expected as a simplified space group of the manganite, $ Pmma $, already allows 21 Raman-active modes\cite{abrashev2001} (the more realistic $ P21/m $ symmetry – that include the MnO$ _{6} $ octahedral tilts - having 54 Raman-active phonon modes). The manganite spectra are also consistent with a disordered rhombohedral phase with $ R\bar{c}3 $ space group \cite{iliev2003}. YBCO has 5 main Raman active phonon modes, with additional modes in special cases of charge- and oxygen-ordering\cite{bakr2013}. An analysis of this spectral region will be presented elsewhere, as here we focus on the Raman scattering below $ 30 $~cm$ ^{-1} $.

In particular, we focus on two prominent features in the low-energy spectra which are marked by the arrow in panel (a) and highlighted in \fig~\ref{fig:raw}(b). In panel (b), both the Stokes and the anti-Stokes signals at $ T=300 $~K in a narrow spectral region around the elastic-line are shown. The modes are clearly pronounced for both the $ z(XX)\underline{z} $ and $ z(X'X')\underline{z} $ geometries, but they are not observed for the cross-polarised geometries which indicates an $ A_{g} $ type symmetry. %The peaks were observed in Raman measurements where the incident beam is 30$ ^{\circ} $ away from the normal to the film surface. 
Importantly, these sharp modes are only observed from superlattice samples. For example, we include in \fig~\ref{fig:raw}(b) a spectrum from a trilayer of the same material, PLCMO(20nm)/YBCO(7nm)/PLCMO(20nm), for which the sharp low-energy features are absent.

To quantitatively characterize these peaks, we fit them using a quadratic background and a pseduo-Voigt line shape as detailed in the supplementary material. An exemplary fit is shown by the magenta line superimposed to the $ z(X'X')\underline{z} $ superlattice data in \fig~\ref{fig:raw}(b), whereas the modelled background is a lighter coloured thin magenta line.
The fitted peaks shown in \fig~\ref{fig:raw}(b) are centred at $ \centre = -12.9 $ and $ -15.2 $~cm$ ^{-1} $, with the absolute uncertainty in the peak positions estimated to be $ 0.3 $~cm$ ^{-1} $, primarily due to systematic uncertainties. The half widths at half maximum (HWHM) are $ \width = 1.0 $ and $ 0.5 $~cm$ ^{-1} $ respectively.
The area of each mode, $ \SW $, is proportional to its Raman susceptibility. While we cannot quantify the Raman susceptibility from our data in absolute terms, we can compare the Bose-corrected areas of these new modes with that of a regular phonon mode. In particular, we find an area of $ \SW \approx 0.2 $~(a.u.) for the phonon excitation at $ \centre \approx 145 $~cm$ ^{-1} $, and areas of $ \SW = 0.5 $ and $ 0.3 $ for the two low-energy modes respectively. This illustrates that the new modes have Raman cross-sections comparable to the weaker phonon modes above 70~cm$ ^{-1} $. % And somewhat larger than the ratios of between $ 0.01 $ to $ 0.13 $  observed in GaAs-AlAs superlattices colvard 1985 PRB out-of-resonance (in resonance, can become comparable).

% Non-linear effect on intensity of signal with laser power, but we data not systematic enough to say much more.

In transition metal-oxides like those studied here, modes in this frequency range might be ascribed to magnetic excitations \cite{murugavel2000}. %[and/Or possibly Phys. Rev. B 95, 174413].
However, for several reasons this is unlikely in our case, despite the significant Mn magnetic moments. Firstly, long-range magnetic order is established only below $ T\approx140 $~K in our PLCMO and NCSMO samples\cite{mallett2016pyp, perret2018, khmaladze2019}, whereas these peaks are intense and sharp already at room-temperature. Secondly, the new modes are only observed for superlattice samples and not in films of the pure manganite material with comparable thickness to the superlattices. Thirdly, whereas magnons are usually observed in crossed polarization, the $ XY $ and $ X'Y' $ polarization channels of our superlattices do not exhibit low-energy modes (\fig~\ref{fig:raw}(b)).  In addition, the low energy modes are not observed in trilayer samples, which rules out that they originate solely due to an interaction between the cuprate and manganite.

\begin{figure}
	\includegraphics[height=5.5cm]{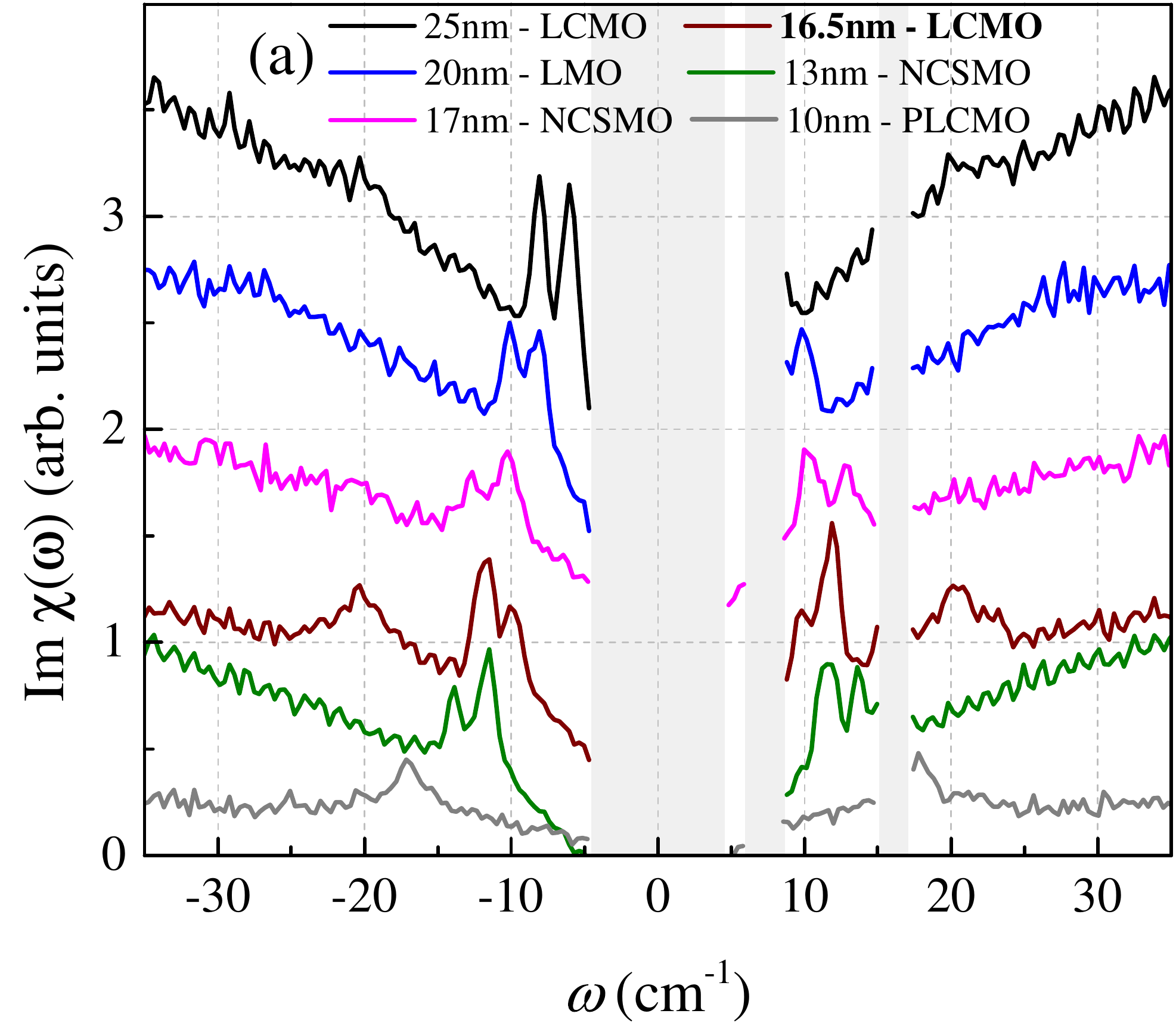}
	\includegraphics[height=5cm]{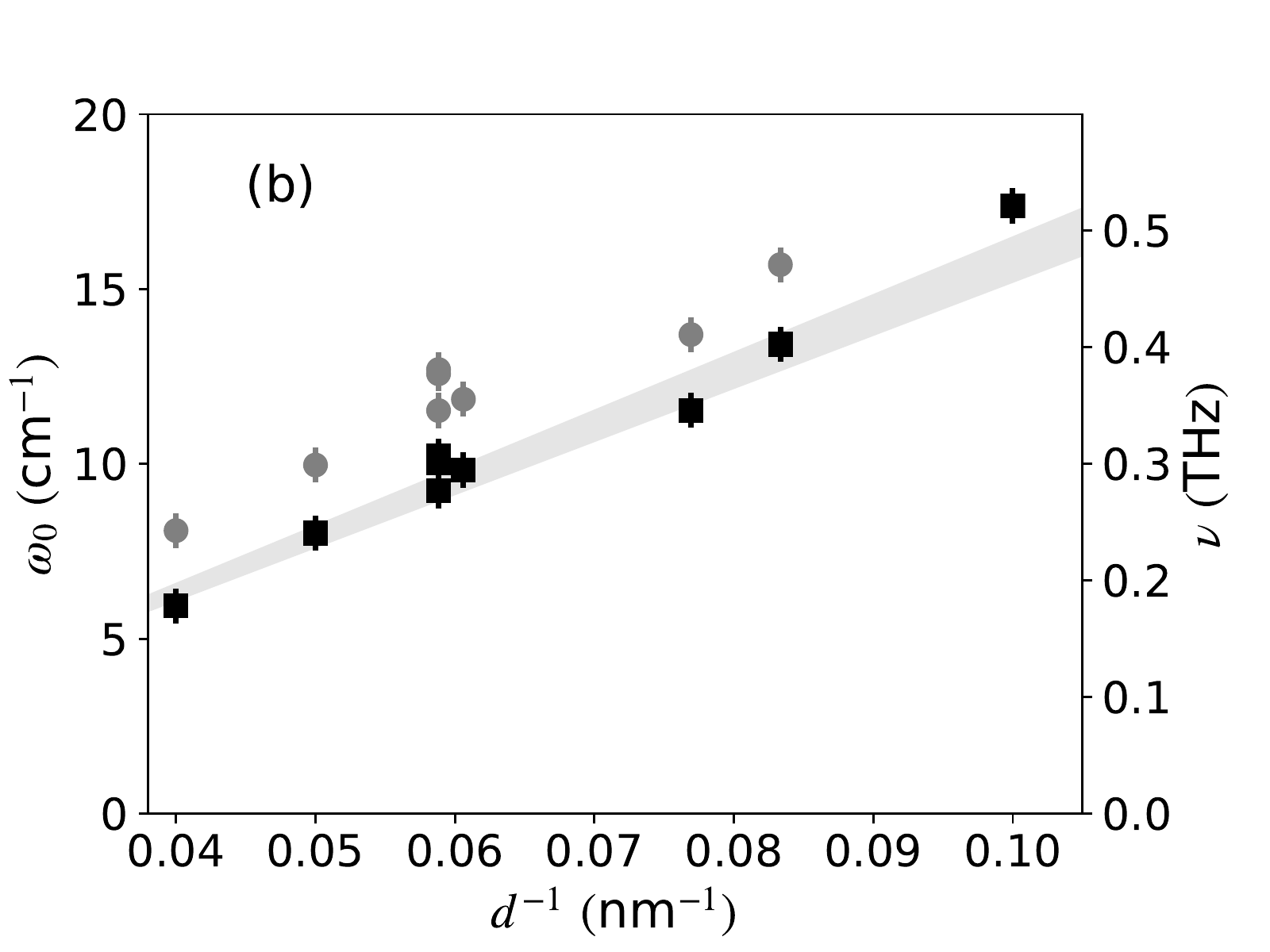}
	\caption{(top) Spectra from multiple superlattice samples with bilayer thickness, $ d $, and the manganite compositions indicated in the legend. (bottom) The centre positions of the two observable low-energy modes versus $ d^{-1} $. The shaded line represents the expected average position of the two low-frequency modes based on the speed of sound in the superlattice.}
	\label{fig:compdep}
\end{figure}

%\subsection{Sample dependence}
Furthermore, \fig~\ref{fig:compdep} illustrates that the position of these peaks depends on the bilayer thickness in the superlattice samples ($i.e.$ the sum of the YBCO and manganite film thickness). \fig~\ref{fig:compdep}(a) shows $ z(X'X')\underline{z} $ spectra for superlattice samples with a range of bilayer thickness, $ d $. Spectra have been offset for clarity. The shift of the modes to lower energy with larger bilayer thickness is clearly seen from the raw data. Figure~\ref{fig:compdep}(b) shows the fitted energies of the two observable low-energy modes, $ \centre $, as data points plotted against $ d^{-1} $.
The solid line in the figure has the form $ \centre = v_s d^{-1} $, where the meaning and choice of the value $ v_s $ is discussed below. % in the Figure are linear fits to the data, and show a good fit over this range which indicates $ \centre \sim d^{-1} $.
The peak areas, $ \SW $, and widths, $ \gamma $, do not appear to show any systematic variation across the samples studied. Finally, we note the possible presence of additional low-energy peaks barely resolved by our measurements, except in particular cases such as the $d= 16.5 $~nm sample (brown curve).

%% Sample dependence %%

These observations show that the low-energy modes most probably arise from a back-folding of the Brillouin-zone, due to the superlattice periodicity, which brings new Raman-active excitations at low energies onto the $ \Gamma $ point.

\begin{figure}
	\includegraphics[width=0.8\columnwidth]{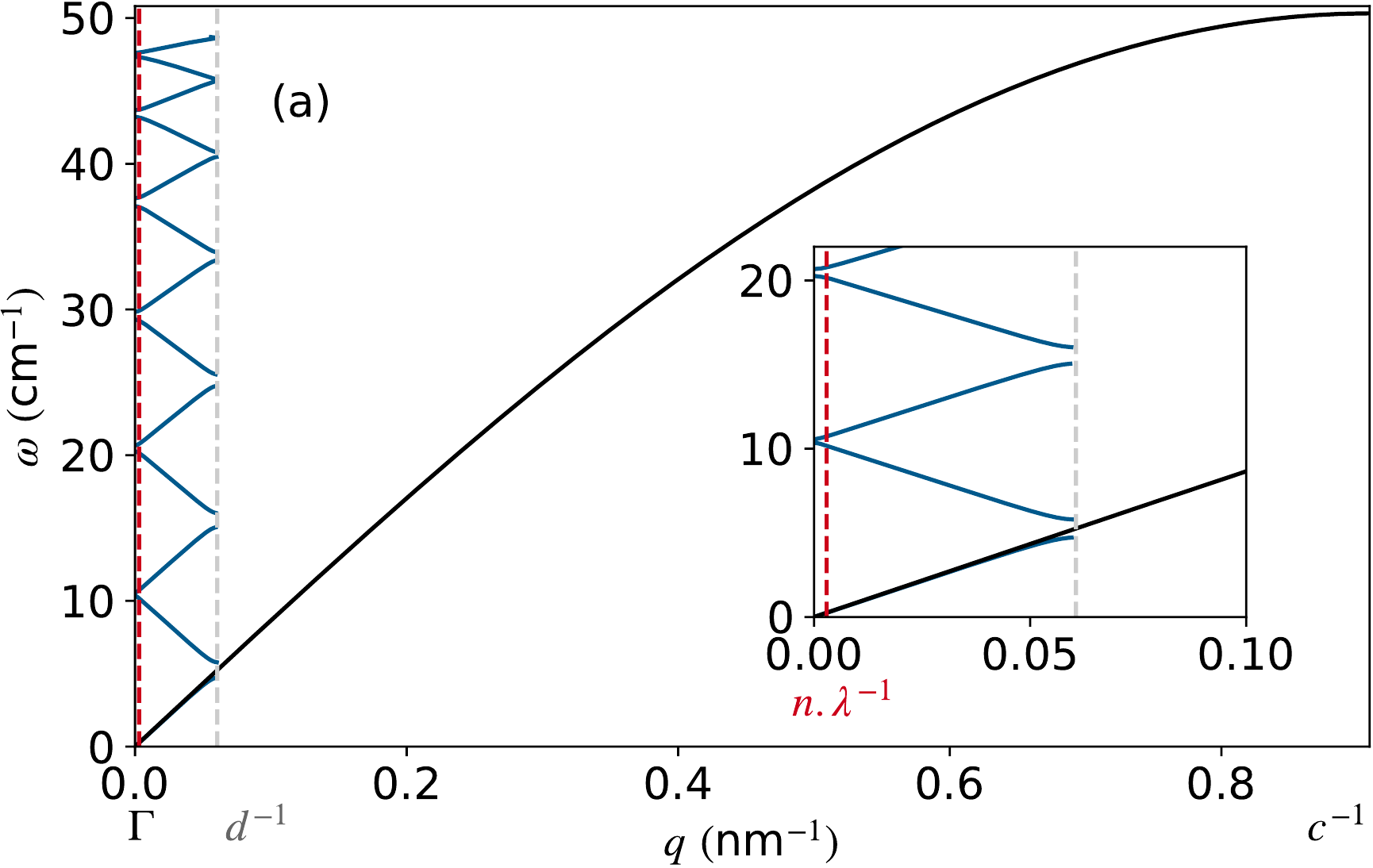}
	\includegraphics[width=0.7\columnwidth]{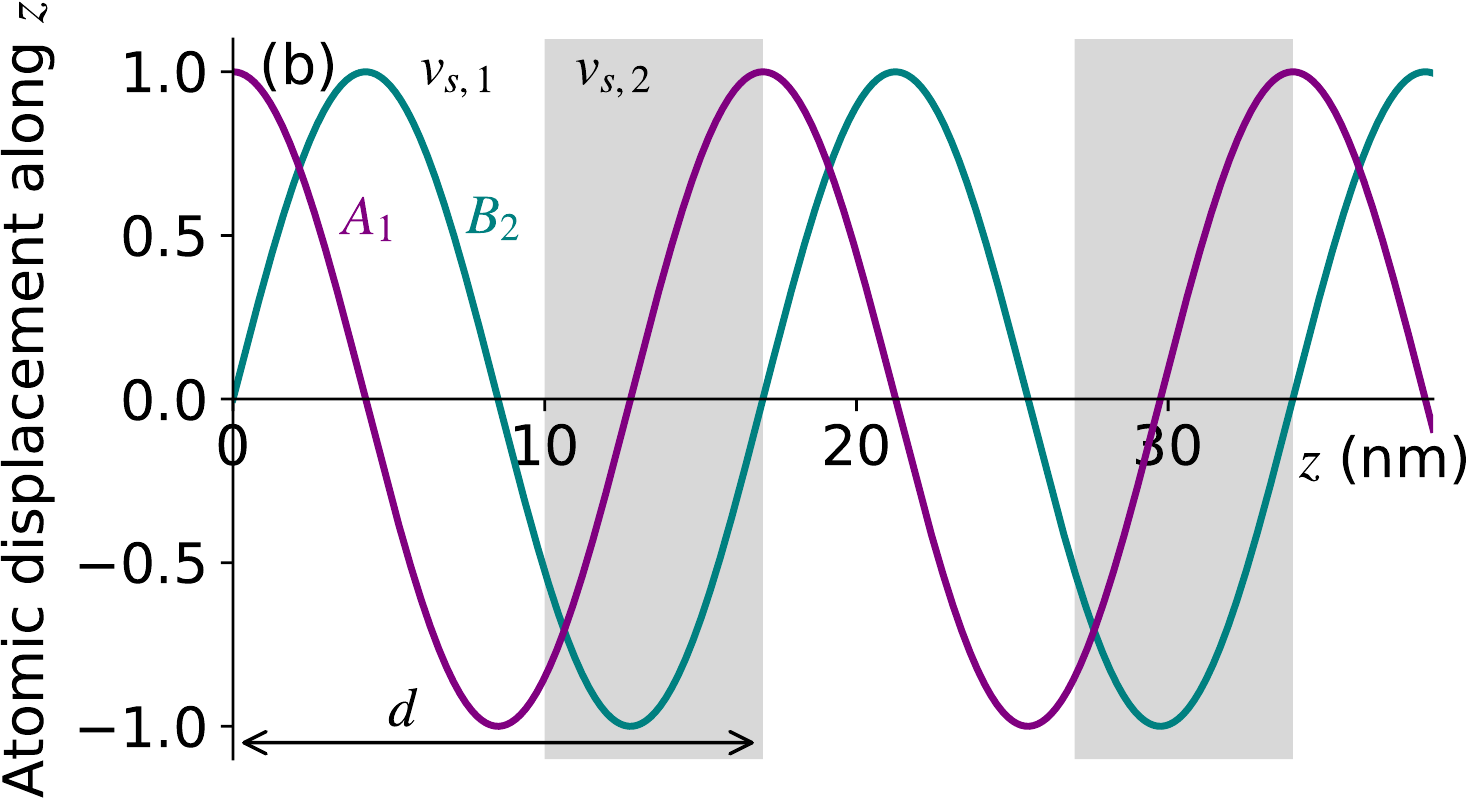}
	\caption{(a) A sketch of phonon backfolding with approximate values for our samples. The black line shows a typical dispersion of an acoustic phonon branch. The additional periodicity of the superlattice, $ d $, backfolds the phonon branch around $ d^{-1} $. There is a splitting of the phonon branches near $ q=0$ and $d^{-1} $. The Raman experiment probes the phonon modes with $ q=n.\lambda^{-1} $, shown with the red dashed line. (b) A sketch of the amplitude of atomic displacement for the two first backfolded modes at $ \Gamma $. The shaded regions represent the material with the higher speed of sound, $ v_{s,2} > v_{s,1}  $. }
	\label{fig:sketch}
\end{figure}

Such a situation has been well documented and analysed in superlattices comprising of GaAs-AlAs semiconductors for acoustic phonon branches \cite{colvard1980, cardona1989, rouhani1991, sapriel1991}. There are well-established models for this situation, starting from either a continuum approximation or linear-chain type models \cite{colvard1985, cardona1989}, of which detailed versions have been developed in order to capture finite-size effects of the sample \cite{dharma-wardana1993}. All such models agree however with the general behaviour captured by the simpler Rytov model \cite{rytov1956}. Within this model, the phonon dispersion is described by
\begin{eqnarray}
\cos(qd) = &&\cos \left( \frac{ \omega d_{1} }{ v_{s,1} }  \right) \cos \left( \frac{ \omega d_{2} }{ v_{s,2} }  \right) \nonumber \\ &&- \frac{ 1 + \kappa^{2} }{ 2\kappa } \sin \left( \frac{ \omega d_{1} }{ v_{s,1} }  \right) \sin \left( \frac{ \omega d_{2} }{ v_{s,2} }  \right) \nonumber
\end{eqnarray}

\noindent where the subscripts $ 1 $ and $ 2 $ denote the two materials in the superlattice and $ v_{s} = \sqrt{ c_{33} / \rho } $ is the sound-velocity (with $ \rho $ the density of the material and $ c_{33} $ is the elastic modulus along the $ c $-axis direction%\footnote{We focus here on the backfolding of the longitudinal acoustic branch as its Raman cross-section is much larger than for the transverse acoustic branches, see e.g. Ref.~\cite{colvard1985}}
). $ d_{1} $ and $ d_{2} $ are the thicknesses of the layers comprising the superlattice and $ \kappa \equiv v_{s,2}\rho_{2}/v_{s,1}\rho_{1} $. This expression describes a folding of the phonon dispersions about $ d^{-1} $, where $ d = d_{1}+d_{2} $, and the opening of gaps at the zone center and boundaries. Figure~\ref{fig:sketch}(a) sketches the back-folding effect with approximate values for our experiments. With the Raman measurements, we (de)excite the phonon modes at $ q=n.\lambda^{-1} $ where $ n $ is the refractive index of the superlattice and $ \lambda $ the laser wavelength.

This model describes accurately our observations. In particular, the straight line in \fig~\ref{fig:compdep}(b) is given by $ v_s d^{-1} $ where $ d $ is obtained from the nominal layer thicknesses (estimated from x-ray reflectivity measurements) and $ v_s = \frac{7}{17} v_{s,\textrm{YBCO}} + \frac{10}{17} v_{\textrm{manganite}} $ is the weighted average of the $ c $-axis speed of sound in the two materials (here we are using the most common 7~nm YBCO layer and 10~nm manganite layer thicknesses) \cite{cardona1989}. The value of $ v_s $ is $ 4750 \pm 200 $~m.s$ ^{-1} $ as determined from the bulk-moduli and densities \cite{hazama2000, jorgensen1990}, but similar values are obtained from other measurements of the speed of sound\cite{seikh2003, li2016, pintschovius1998, weber2011neutrons}.%\cite{almond1988, baumgart1989, seikh2003, ren2006, li2016, pintschovius1998, weber2011neutrons}.
The width of the line in \fig~\ref{fig:compdep}(b) comes from the uncertainty in %$ v_{s,\textrm{YBCO}} $ and $ v_{s,\textrm{manganite}} $.
$ v_{s} $ for the individual YBCO and manganite layers.
$ v_s d^{-1} $ represents an average of the two peak-positions which are split due to mixing and the finite-$ q $ of the laser line\cite{colvard1985}. Therefore, our experimental results are consistent with a slightly higher value of $ v_{s} \approx 5150 $~m.s$ ^{-1} $.

A generalized relaxation time of the mode can be expressed as $ \tau = \gamma^{-1} $, which is approximately 40~ps. If we assume $ v_{s}\sim 5000 $~m.s$ ^{-1} $, then the scattering length of the mode is $ l = v_{s}\tau \approx 200 $~nm. This is close to the total film thickness and suggestive of scattering of the phonon at the top surface of the superlattice and bottom interface with the substrate.

We now discuss the intensity of the peaks.
Close to the Brillouin zone center, the upper and lower branches develop preferential $ A_{1} $ or $ B_{2} $ symmetry. The atomic displacement amplitude for the two symmetries are sketched in \fig~\ref{fig:sketch}(b), following Ref.~\cite{colvard1980}. The $ B_{2} $ mode is not Raman-active (off-resonance), and therefore the relative Raman cross-section of the two branches will be different near the $ \Gamma $ point. The relative thickness and speed of sound of the manganite and YBCO layers determine whether the upper or lower branch has dominant $ B_{2} $ symmetry. We denote these as $ d_1 $ and $ d_2 $, respectively, whereby layer 1 (the manganite) has a slower speed of sound than layer 2 (YBCO): $ v_{s,1} < v_{s,2} $. For $ d_{1} < d_{2} $, the $ B_{2} $ mode has lower energy than the $ A_{1} $ mode which results in a smaller intensity of the lower energy mode in the Raman spectrum. The sample representing this case is marked in bold in the legend of \fig~\ref{fig:compdep}(a). For most of our samples $ d_{1} > d_{2} $ holds, so that the situation is reversed with the higher energy mode having a lower intensity. For $ d_{1} = d_{2} $, the second-order back-folded phonon branch will not be Raman-active. This is the case for the ``20nm - LMO'' sample in \fig~\ref{fig:compdep}(a) and approximately the case for the samples with 10~nm and 7~nm layer thickness, which may be the reason we only rarely see the second-order back-folded modes.
More quantitative predictions of the phonon intensities require first-principles calculations of the photoelastic coefficients \cite{colvard1985}.
% with a ..... PHOTOELASTIC MODEL ..... . This framework satisfactorally described intensities in GaAs-AlAs superlattices, off-resonance and away from the zone-edge.

%The upper and lower branches in \fig~\ref{fig:sketch}, which give rise to the two peaks in the Raman spectra, have mixed $ A_{1} $ and $ B_{2} $ symmetry character, the latter of which is not a Raman-active symmetry (off-resonance). The degree of admixture is governed by their proximity to the Brillouin zone centre or edge, and the relative dominance of $ A_{1} $ or $ B_{2} $ depends on the relative thickness of the manganite and ybco layers.
%This can lead to unequal relative intensities of the peaks in the doublet if the branches have different admixtures. The mixing is in turn dependent on the relative thickness...

%Coherent acoustic phonons have been observed by fs pump-probe reflectivity studies of Li \textit{et al.} \cite{li2016} in 3 repetition superlattices of LCMO (10nm) / YBCO (10,5nm) grown on STO. However, they find good agreement between their observed acoustic phonon frequency in the superlattice, and that expected from the previously reported sound velocities for bulk materials.

%Neutron studies \cite{pintschovius1998}, and from phonon-d	ispersions calculated from first-principles \cite{, rini2007} indicate however $ v_{s} \sim 350 $~m.s$ ^{-1} $??.

We also note that we observe similar low-frequency modes in superlattices grown in our laboratory where the YBCO is replaced for another metal oxide. These include SrFeO$ _{3} $/La$ _{2/3} $Ca$ _{1/3} $MnO$ _{3} $ \cite{perret2017} and SrRuO$ _{3} $/La$ _{2/3} $Ca$ _{1/3} $MnO$ _{3} $ superlattices. % and the data are presented in the supplementary information. 

%\subsection{Spatial dependence}

The results above show that these low-energy modes can be used to determine the spatial variation in a superlattice's bilayer thickness (or more generally its repeat-unit thickness), $ d $, because (i) their position, $ \centre $, is a function of $ d $ and (ii) the micro-Raman technique we use here has a 10~$ \mu $m$ ^{2} $ spatial resolution in the plane. Spatial uniformity is of particular interest for growth techniques whereby the deposition rate of the films may not be constant across the sample. One particularly important example in terms of oxide materials is the growth by PLD onto substrates with surface area comparable to the size of the plasma plume. The spatial dependence of the intensity and HWHM may also be used to characterize film quality.

\begin{figure}
	\includegraphics[height=6cm]{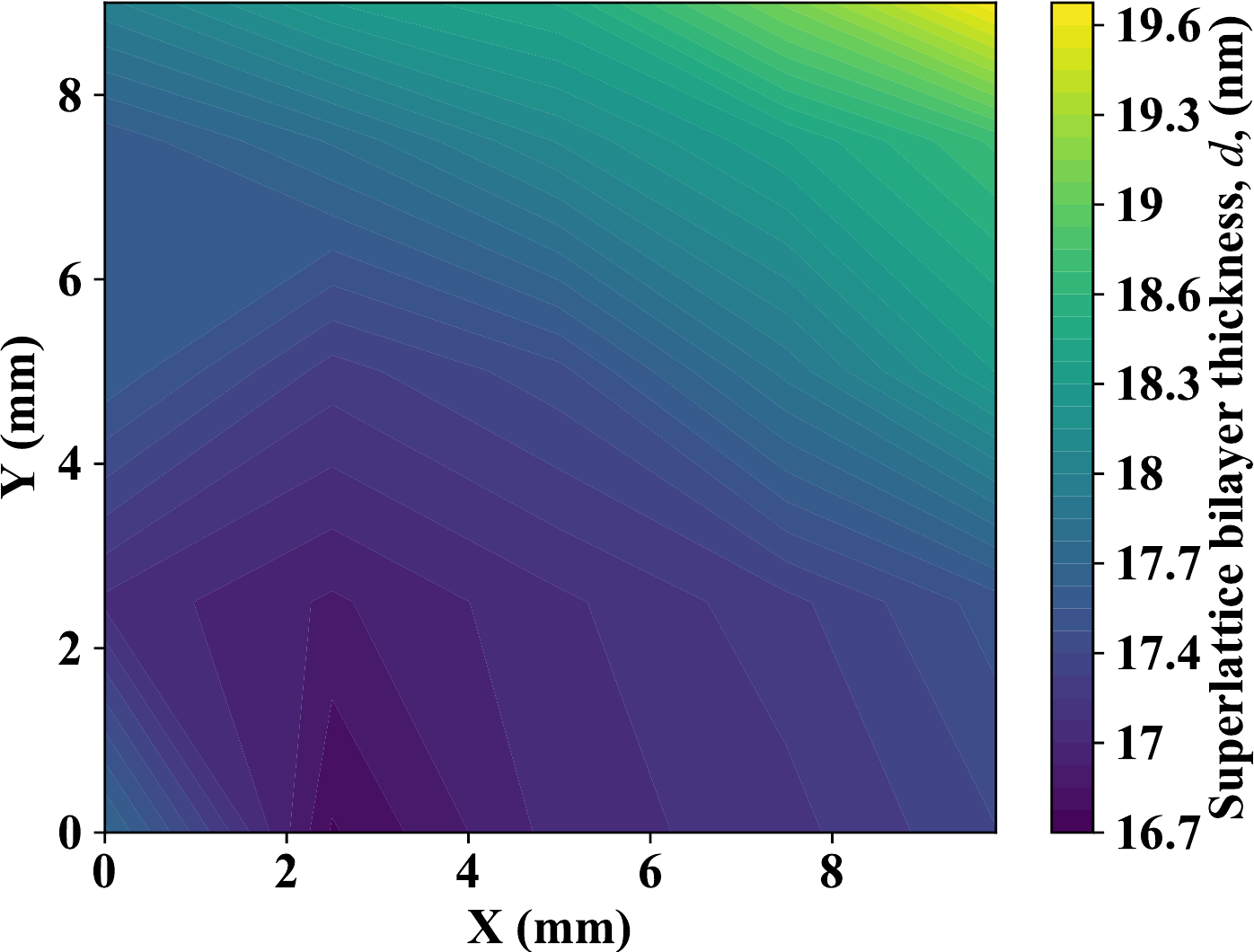}
	\caption{A contour plot of the combined thickness of the LCMO and YBCO layers, $ d $, determined from $ \centre $ of the low-frequncy modes. There is smooth gradient in $ d $ from one corner to the opposite.
	}
	\label{fig:SpatialMap}
\end{figure}

To exemplify the potential of our method, we performed measurements on a 7x7 points grid across the surface of a nominally LCMO(10 nm)/YBCO(10 nm) superlattice grown by PLD on an LSAT with surface area $10 \times 10$~mm$ ^{2} $. The results are summarized in Fig.~\ref{fig:SpatialMap}. The colour scale represents $ d $ as determined from the fitted $ \centre $ of the two low-energy modes at each point and $ v_{s} = 5000 $~m.s$ ^{-1} $ as determined in Fig.~\ref{fig:compdep}(b). The data are linearly interpolated between the measurement points and we estimate the uncertainty in $ d $ to be $ \sim 1 $~nm. Fig.~\ref{fig:SpatialMap} shows that the LCMO-YBCO layer thicknesses are not uniform across the approximate $9 \times 9$~mm$ ^{2} $ area that was measured. Instead, we identify a smooth gradient in the value of $ d $ from one corner to the opposite corner.

%\subsection{Temperature dependence}
%% Temperature dependence %%

Finally, we note that we observed only a subtle temperature dependence of the low-frequency modes between room-temperature to 10~K, as shown in the supplementary information. This is despite the various electronic and magnetic phase transitions that our samples exhibit below 300~K \cite{ mallett2016pyp, perret2018, khmaladze2019}.%{golod2013, sen2016}.
% Hence, higher resolution data would be needed in order to study the coupling between this phonon-branch and the spin or low-energy electronic systems. 

%\section{Discussion and Conclusions}

%There are several ways in which future measurements could be improved to gain a richer characterization of the superlattice film quality. Firstly, higher-resolution spectra would better resolve the peak positions, intensity, widths and asymmetry. Such higher resolution spectra would be required to resolve subtle structural transitions or coupling between the phonon modes and the various magnetic and electronic phase transitions in these materials at lower temperatures. Furthermore, it may be possible to observe fine structure of these peaks which would allow an analysis beyond the continuum Rytov model, for example using finite-size, linear-chain type models that can model thickness variations in the superlattice \cite{dharma-wardana1993}. Secondly, it would be possible to independently determine both $ d $ and $ v_{s} $ from the Raman data alone, with measurements using multiple laser wavelengths. Such measurements would also reveal in more detail the splitting of the two phonon branches that is caused by inter-atomic coupling between the two metal-oxide layers \cite{colvard1985}.

In summary, we report systematic measurements of low-frequency modes in metal-oxide superlattices grown by pulsed laser deposition. We show that these modes arise from a back-folded $ c $-axis acoustic phonon branch. As such, their observation demonstrates the high quality of the thin-film superlattices. The modes can be used to characterize the bilayer thickness of the superlattice and/or the $ c $-axis speed of sound. We utilized the spatial resolution of the Raman microscopy to map the film thickness inhomogeneity across a larger $ 10 \times 10 $~mm sample. This information is important for monitoring and improving the quality of future metal-oxide superlattices which might comprise the elementary building blocks of next generation electronic devices.

%\section{Supplementary material}
%See supplementary materials for details of the methods and samples, as well as data showing the temperature dependence of the backfolded modes. % and the low-frequency modes in alternate metal-oxide superlattices. 

\begin{acknowledgments}
This work was supported by the Schweizerische Nationalfonds (SNF) through grant No. 200020-172611. BPPM acknowledges support from the Rutherford Foundation of New Zealand. AC acknowledges support of the MacDiarmid Institute. We thank Dr. F. Weber and Dr. R Heid for valuable discussions on this work.
\end{acknowledgments}

%\bibliography{C:/Users/Ben/Documents/work/papers_me/thesis/literature}


\end{document}